\documentclass[12pt, letterpaper]{article}
\usepackage{amsmath, amsfonts, amsthm, amssymb, anysize, graphicx, fullpage, setspace, hyperref, thmtools, mathrsfs}
\usepackage[authoryear, round]{natbib}
\usepackage{enumerate}
\usepackage{float}
\restylefloat{table}

\declaretheorem[style=definition,numberwithin=section]{example}

\newcommand{\var}{\mathrm{var}}

\title{Measuring the Initial Transient: Reflected\\Brownian Motion}
\author{Rob J. Wang\thanks{Department of Management Science and Engineering, Stanford University.
Stanford, CA, 94305. Email Address: robjwang@stanford.edu.}\hspace{1 cm}
Peter W. Glynn\thanks{Department of Management Science and Engineering, Stanford University.
Stanford, CA, 94305. Email Address: glynn@stanford.edu.}}
\begin{document}
\date{May 27, 2014}

\maketitle
\renewcommand{\baselinestretch}{1}  
\numberwithin{equation}{section}

  \begin{abstract}
	We analyze the convergence to equilibrium of one-dimensional reflected
	Brownian motion (RBM) and compute a number of related initial transient formulae.
	These formulae are of interest as approximations to the initial transient
	for queueing systems in heavy traffic, and help us to identify settings in which
	initialization bias is significant. We conclude with a discussion of mean square
	error for RBM. Our analysis supports the view that initial transient effects
	for RBM and related models are typically of modest size relative to the intrinsic
	stochastic variability,	unless one chooses an especially poor initialization.
	\end{abstract}

  \section{Introduction}
	
	This paper is concerned with using one-dimensional reflected Brownian motion (RBM)
	as a theoretical vehicle for studying the initial transient problem. Given that
	RBM is a commonly used approximation to a wide variety of different queueing models,
	the initial transient behavior of RBM can be viewed as being representative of a large
	class of simulation models in which congestion is a key factor.
	
	The stochastic process $X = (X(t):\,t \geq 0)$ is said to be a (one-dimensional) RBM
	if it satisfies the stochastic differential equation (SDE)
	\[dX(t) = -r dt + \sigma dB(t) + dL(t),\]
	where $B = (B(t):\,t \geq 0)$ is standard Brownian motion and $L$ is a continuous
	nondecreasing process that increases only when $X$ is at the origin
	(so that $I(X(t) > 0)dL(t)=0)$.
	In particular, the process $L$ is a ``boundary process" that serves to keep $X$ nonnegative
	as befits an approximation to a queue. The parameter $-r$ represents the ``drift"
	of the RBM, and $\sigma$ is its ``volatility" parameter.
	
	To illustrate the sense in which RBM can be used to approximate a queue, consider a system with
	a single queue that is being fed by a renewal arrival process in which $\chi$ denotes a 
	generic interarrival time random variable (rv). Customers are served by one of
	$m$ identical servers, in the order in which they arrive. The service times are
	independent and identically distributed (iid) across the servers and across the customers,
	and are also independent of the interarrival times. If $V$ is a rv having the common service
	time distribution, set $\lambda = 1/E\chi,\;\mu = 1/EV$
	and $\sigma_A^2 = \var\,\chi,\; \sigma_S^2 = \var\, V$.
	It is well known that if $Z(t)$ is the number-in-system at time $t$, then
	\[Z(\cdot) \stackrel{\mathscr D}{\approx} X(\cdot),\]
	where $X$ is an RBM with $r = m \mu - \lambda$ and $\sigma^2 = \lambda^3 \sigma_A^2 + m \mu^3 \sigma_S^2$,
	provided that $m\mu-\lambda$ is small (so that the system is in ``heavy traffic")
	and $t$ is of the order $(m\mu-\lambda)^{-2}$. Here, $\stackrel{\mathscr D}{\approx}$
	means ``has approximately the same distribution as", and the rigorous support rests on a so-called
	``heavy traffic" limit theorem; see, for example, \cite{Iglehart}.
	
	This paper begins with a discussion of the convergence to equilibrium of RBM (Section 2).
	It subsequently develops closed-form expressions for various initial transient quantities
	associated with RBM, distinguishing between ``functional" (Section 3) and ``distributional" (Section 4)
	perspectives. These expressions can be used to help plan steady-state/equilibrium
	simulations of queueing models that can be approximated by RBM as well as to identify settings
	in which initial bias is significant (Section 5). We conclude by deriving a decomposition of
	mean square error (MSE) in the setting of RBM (Section 6).
	
	Though only in the context of RBM, the results in this paper are intended develop general insights
	into the initial transient problem that can be of potentially broader applicability.

	\section{Convergence to Equilibrium for RBM}
	
	It is well-known that if $r > 0$, then $X$ has an equilibrium, in the sense that
	\[X(t) \Rightarrow X(\infty)\]
	as $t \rightarrow \infty$, where $\Rightarrow$ denotes weak convergence. The distribution
	of $X(\infty)$ is given by
	$P(X(\infty) \in dx) \stackrel{\Delta}{=} \pi(dx)
	= \eta e^{-\eta x}dx$
	for $x \geq 0$, where $\eta \stackrel{\Delta}{=}\frac{2r}{\sigma^2}$
	(see, for example, \cite{Harrison} p.94). A key question in
	the study of the initial	transient problem for $X$ is its rate of
	convergence to equilibrium. One vehicle for
	studying this question is the well-known formula
	\begin{equation}\label{transitionDensity}
	p(t,\,x,\,y) = \eta e^{-\eta y} \Phi\left(\frac{r  t-x-y}{\sigma \sqrt{t} }\right)
	+\frac{1}{\sigma \sqrt{t}}\phi\left(\frac{-r  t+x-y}{\sigma \sqrt{t}}\right)
	+ \frac{1}{\sigma \sqrt{t}} e^{-\eta y}	\phi\left(\frac{-r t+x+y}{\sigma \sqrt{t}}\right)
	\end{equation}
	for the transition density of $X$; see \cite{Harrison} p.49. Here,
	$p(t,\,x,\,y)dy \stackrel{\Delta}{=} P(X(t) \in dy\,|\,X(0)=x)$, $\Phi(x) = P(N(0,\,1) \leq x)$
	(where $N(0,\,1)$ denotes a normal rv with mean $0$ and unit variance),
	and $\phi(x)$ is the density associated with $\phi$. But it is difficult to
	``read off" the rate of convergence from (\ref{transitionDensity}).
	
	However, an alternative representation for the transition density of $X$ can be computed.
	Recall that the rate at which the transition probabilities of a Markov jump process converge to
	their respective equilibrium probabilities can easily be determined once the eigenvalues
	and eigenvectors of the rate matrix are known. Something similar can be implemented in the RBM
	setting. This leads to an alternative representation of the transition density known as the
	\textit{spectral decomposition}.
	
	To begin, note that It\^o's formula	gives
	\begin{eqnarray*}
	d(e^{-\lambda t} u(X(t)))
	&=& -\lambda e^{-\lambda t} u(X(t))dt + e^{-\lambda t} u'(X(t))dX(t) + e^{-\lambda t} \frac{u''(X(t))}{2} \sigma^2 dt\\
	&=& e^{-\lambda t} ((\mathscr L u)(X(t)) - \lambda u(X(t)))dt + e^{-\lambda t}
	u'(X(t))\sigma dB(t) + e^{-\lambda t} u'(X(t))dL(t),
	\end{eqnarray*}
	where $\mathscr L$ is the second order differential operator given by
	\[\mathscr L \stackrel{\Delta}{=} -r \frac{d}{dx} + \frac{\sigma^2}{2} \frac{d^2}{dx^2}.\]
	Because $L$ increases only when $X$ is at the origin, $u'(X(t))dL(t) = u'(0)dL(t)$.
	Consequently, if $u'(0) = 0$ and the stochastic integral is integrable,
	\[e^{-\lambda t}u(X(t)) - \int_0^t e^{-\lambda s} ((\mathscr L u)(X(s)) - \lambda u(X(s)))ds\]
	is a martingale. It follows that if $u$ also satisfies
	\begin{equation}\label{SturmLiouville}
	\mathscr L u = \lambda u
	\end{equation}
	(subject to $u'(0)=0$), then
	$E_x u(X(t)) = e^{\lambda t} u(x)$ for $t,\,x \geq 0$, where
	$E_x(\cdot) \stackrel{\Delta}{=} E(\cdot\,|\,X(0)=x)$.
	If $\lambda < 0$, sending $t\rightarrow \infty$
	allows us	to conclude that $Eu(X(\infty)) = 0$, and the rate of convergence
	of $E_xu(X(t))$ is exponentially fast with associated rate parameter $\lambda$.\\
	
	\noindent
	\textit{Remark:} This makes clear that the rate of convergence of $E_xf(X(t))$ to $Ef(X(\infty))$
	often depends on the choice of $f$.\\
	
	For each $\lambda \in \mathbb R$, there exists a nontrivial solution (unique up to multiplicative constants)
	to (\ref{SturmLiouville}), which can be found by direct computation. However, according to \cite{Linetsky},
	the \textit{spectrum} (in the operator theoretic sense) consists only of
	$\lambda = 0$ and	$\lambda \leq -\gamma$, where $\gamma \stackrel{\Delta}{=} r^2/(2\sigma^2)$.
	As a consequence, the distance	between the top two points in the spectrum, namely $0$ and $-\gamma$,
	is equal to $\gamma$.	This distance is known, in the Markov process literature,
	as the \textit{spectral gap} of $X$.
	
	For	$\lambda \in (-\infty,-\gamma)$, let
	\[u_\lambda(x) \stackrel{\Delta}{=} e^{\frac{r x}{\sigma^2}}
	\left( s(\lambda) \cos\left(\frac{s(\lambda)x}{\sigma^2}\right)
	-r \sin\left(\frac{s(\lambda)x}{\sigma^2}\right)\right).\]
	where $s(\lambda) = \sigma \sqrt{-2(\lambda +\gamma)}$. 
  There is a standard ``recipe" for constructing the transition density
	for reversible diffusion processes (of which RBM is one)
	that can be found, for example, on p.332 of \cite{Karlin}. When specialized to the RBM setting,
	one obtains the spectral decomposition for $p(t,\,x,\,y)$, namely
	\begin{eqnarray*}
	p(t,\,x,\,y)
	&=& \frac{2r}{\sigma^2}e^{-\frac{2r}{\sigma^2}y} + \frac{2}{\pi\sigma^2} e^{-\frac{r(y-x)}{\sigma^2}-\frac{r^2t}{2\sigma^2}}\\
	&\qquad\cdot& \int_0^\infty
	\frac{e^{-\frac{v^2 t}{2\sigma^2}}}{v^2+r^2} \left(v\cos\left(\frac{vx}{\sigma^2}\right)-r\sin\left(\frac{vx}{\sigma^2}\right)\right)
	\left(v\cos\left(\frac{vy}{\sigma^2}\right)-r\sin\left(\frac{vy}{\sigma^2}\right)\right)dv\\
	&=& \eta e^{-\eta y} - \frac{1}{2\pi r}	\cdot \int_{-\infty}^{-\gamma} e^{\lambda t} u_\lambda(x)u_\lambda(y) \frac{1}{s(\lambda) \lambda}
	\eta e^{-\eta y} d\lambda
	\end{eqnarray*}
	(see \cite{Linetsky}). We thus find that, for $f$ appropriately integrable,
	\begin{equation}\label{expectation}
	E_xf(X(t)) = Ef(X(\infty))  - \frac{1}{2\pi r}	\cdot \int_{-\infty}^{-\gamma}
	e^{\lambda t} u_\lambda(x) \langle f,\,u_\lambda \rangle\frac{1}{s(\lambda) \lambda}d\lambda,
	\end{equation}
	where
	\[\langle f,\,u_\lambda \rangle \stackrel{\Delta}{=}
	\int_0^\infty f(y)u_\lambda(y) \eta e^{-\eta y} dy.\]
	
	The spectral representation (\ref{expectation}) makes clear that the spectral gap $\gamma$
	is precisely the exponential rate constant governing the rate at which
	$E_xf(X(t))$ converges to $Ef(X(\infty))$. Furthermore, when $t$ is large, it is primarily
	the ``projection" of $f$ onto $u_{-\gamma}$ (i.e. the magnitude of
	$\langle f,\,u_\lambda \rangle$)
	that determines the magnitude of $E_xf(X(t)) - Ef(X(\infty))$ (given that the integral
	is largely determined by the integrand's contribution from a neighborhood of $\lambda = -\gamma$).

	\section{The Initial Transient Effect}
	
	Given a performance measure $f$, we have studied in Section 2 the rate of convergence of
	$E_xf(X(t))$ to its equilibrium value $Ef(X(\infty))$. Our goal here is to compute the
	magnitude of the initial transient effect, assuming that no deletion
	is implemented. This can inform our decision as to how serious the initial transient
	effect is, and whether/how deletion is warranted. Because the typical estimator
	used to compute an equilibrium quantity in a simulation context is a time-average, the
	effect of the initial transient in the steady-state simulation setting is, in some sense,
	an integrated version of the theory of Section 2.
	
	To compute the effect of the initial transient, note that if $h$ is twice continuously
	differentiable with $h'(0) = 0$, then It\^o's formula yields
	\[dh(X(t)) = (\mathscr L h)(X(t))dt + h'(X(t))\sigma dB(t).\]
	It follows that if the stochastic integral is integrable and $\mathscr L h = -f_c$
	(with $f_c(x) = f(x) - Ef(X(\infty))$ for $x \geq 0$), then
	\[M(t) \stackrel{\Delta}{=} h(X(t)) + \int_0^t f_c(X(s))ds = \int_0^t h'(X(s))\sigma dB(s)\]
	is a martingale, from which we conclude that
	\[\int_0^t E_x f_c(X(s))ds = h(x) - E_x h(X(t)).\]
	In view of our discussion in Section 2, we expect that
	\[E_x h(X(t)) = Eh(X(\infty)) + O(e^{-\gamma t})\]
	as $t \rightarrow \infty$, where $O(a(t))$ represents a function for which $O(a(t))/a(t)$
	remains bounded as $t \rightarrow \infty$. Thus
	\[E_x \frac{1}{t} \int_0^t f(X(s))ds = Ef(X(\infty)) + \frac{1}{t} h_c(x) + O(e^{-\gamma t})\]
	as $t \rightarrow \infty$, where $h_c(x) = h(x) - Eh(X(\infty))$. Hence, the constant $h_c(x)$
	expresses (up to an exponentially small order) the bias of the estimator $t^{-1} \int_0^t f(X(s))ds$ induced by the initial transient.
	
	We now turn to computing $h_c(x)$. The solution $h$ to \textit{Poisson's equation}
	\[(\mathscr L h)(x) = -f_c(x),\;\;x \geq 0\]
	\[h'(0)=0,\;\;h(0) = 0\]
	is given by
	\[h(x) = -\frac{1}{r} \int_0^x f_c(y) (e^{-\eta (y-x)}-1)dy,\]
	from which we can conclude that
	\begin{equation}\label{integral}
	h_c(x) = -\frac{1}{r}\int_0^x f_c(y) \left(e^{-\eta(y-x)}-1\right) dy
	+ \frac{2}{\sigma^2} \int_0^\infty \int_0^x f_c(y)(e^{-\eta y}-e^{-\eta x})dydx.
	\end{equation}
	
	\begin{example}
  For $f(x) = x$,
	\[h_c(x) = \frac{x^2}{2r} - \frac{\sigma^4}{4 r^3}.\]
	\end{example}

	\begin{example}
	For $f(x) = x^2$,
	\[h_c(x) = \frac{x^3}{3r} + \frac{\sigma^2 x^2}{2r^2} - \frac{\sigma^6}{2r^4}.\]
	\end{example}

  \begin{example}
	For	$f(x) = e^{\theta x}$ (with $\theta < \eta$),
	\[h_c(x)=\frac{-\theta ^2 \sigma ^4+4 r^2 \left(-\theta x+e^{\theta  x}-1\right)
	+2 \theta  r \sigma ^2 \left(\theta  x-e^{\theta  x}+1\right)}{\theta  r \left(\theta  \sigma ^2-2 r\right)^2}.\]
	\end{example}

	\begin{example}
	For $f(x) = I(x > b)$, where $I(\cdot)$ is the indicator function,
	\[h_c(x) =
	\begin{cases}
	\frac{e^{-\frac{2 b r}{\sigma ^2}} \left(\sigma ^2 \left(e^{\frac{2 r x}{\sigma ^2}}-1\right)-2 r (b+x)\right)}{2 r^2} & x < b\\
		\frac{e^{-\frac{2 b r}{\sigma ^2}} \left(e^{\frac{2 b r}{\sigma ^2}} \left(-2 b r+2 r x+\sigma ^2\right)-2 r (b+x)-\sigma ^2\right)}{2 r^2} & x \geq b.
	\end{cases}\]
	\end{example}

	Given the performance measure $f$, the functional \textit{unconditional initial transient effect}
	(UNITE) measure is given by
	\[\beta_f(\mu) = \left| \int_0^\infty h_c(x) \mu(dx) \right|\]
	for a given initial distribution $\mu$, while the functional \textit{conditional
	initial transient effect} (CITE) measure is defined by
	\[\tilde{\beta}_f(\mu) = \int_0^\infty \beta_f(\delta_x) \mu(dx),\]
	where $\delta_x(\cdot)$ is a unit point mass distribution at $x$; see \cite{MSERPaper}
	for a more detailed discussion of these measures. Note that in a single replication setting,
	the initial transient effect is determined by the random placement of $X(0)$, so that
	averaging the effect over the initial distribution $\mu$ (as in CITE) seems reasonable.
	Given this viewpoint, it is then appropriate to view $\tilde{\beta}_f(\pi)$ as a benchmark
	against which $\tilde{\beta}_f(\mu)$ for other initializations $\mu$ can be compared.
	(After all, initializing with $\pi$ makes $X$ a stationary process in which no initial
	transient is present.)
	
	In particular, we can now separate the state space into ``good states" and ``bad states",
	depending on whether $\beta_f(\delta_x) \leq c \tilde{\beta}_f(\pi)$ for the state $x$
	or not, where $c$ is a given constant. For example, for $c=1$, the good states are all
	those states $x$ for which $\beta_f(\delta_x)$ is smaller than the CITE measure associated
	with $\pi$. Of course, the set of good states is sensitive to the choice of $c$. Hence, it
	is of interest to study the dependence of the set of good states on the parameter $c$;
	see Figure 1 below. We note that the notion of good state is related to the ``typical state"
	idea introduced by \cite{GrassmannConference} and \cite{GrassmannJournal}.
	
	For $f(x) =x$, the set of good states
  $\mathscr G (r,\,\sigma^2;\,c) \stackrel{\Delta}{=} \{x:\, \beta_f(\delta_x) \leq c \tilde{\beta}_f(\pi)\}$
	is given by
	\begin{eqnarray*}
	\mathscr G (r,\,\sigma^2;\,c)
	&=& \left\{ x:\, 	\left|\frac{x^2}{2r}-\frac{\sigma^4}{4r^3}\right|
	\leq c \left(\frac{ (1+\sqrt{2})e^{-\sqrt{2}}\sigma^4 }{2r^3}\right) \right\}\\
	&=& \left\{ x:\, 	\left|\frac{1}{2}\left(\frac{x}{EX(\infty)}\right)^2-1\right|
	\leq 2(1+\sqrt{2}) c e^{-\sqrt{2}} \right\}.
	\end{eqnarray*}
	For this performance measure, $\mathscr G (r,\,\sigma^2;\,c)
	= EX(\infty)\mathscr G (1,\,2;\,c)$,
	so it suffices to graph
	$\mathscr{G}(r,\,\sigma^2) \stackrel{\Delta}{=}
	\{(x,\,c):\, x \in \mathscr G (r,\,\sigma^2;\,c),\; c \geq 0\}$
	only for $EX(\infty) =1$; the resulting graph can be found below. Note that for $c=1$, the
	set of good states already covers $[0,\,2.09]$, so that the set of initial values 
	providing single-replication bias characteristics roughly comparable to that associated with
	initializing under $\pi$ is quite robust (in fact, more than twice the mean).

	\begin{figure}[H]
	\begin{center}
	\includegraphics[scale = 0.30]{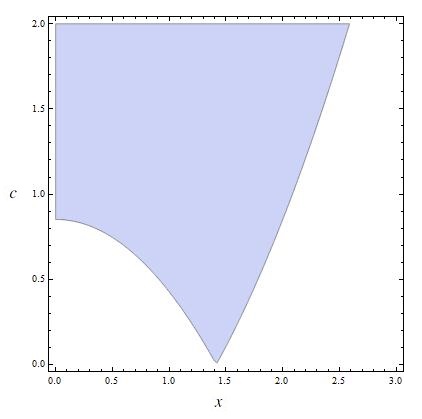}
	\caption{Plot of Good States: $EX(\infty) =1$}
	\end{center}
	\end{figure}

	\section{The Distributional Initial Transient}

	The convergence of $X(t)$ to $X(\infty)$ involves the entire distribution of $X$
	(as opposed to only the convergence of $E_x f(X(t))$ for a single performance measure $f$),
	so that much of the probability literature is concerned with the rate of convergence
	at which the distribution of $X(t)$ approaches that of $X(\infty)$. In fact,
	one commonly used measure for judging the rate of convergence to equilibrium is the
	\textit{weighted total variation norm} defined by
	\[\sup_{|f| \leq w} |E_\mu f(X(t)) - E f(X(\infty))|\]
	for a given weight function $w:\,\mathbb{R}_+ \rightarrow \mathbb{R}_+$.
	
	By analogy, with the discussion of Section 3, it is natural to
	study the distributional UNITE measure given by
	\[\beta(\mu) \stackrel{\Delta}{=} \sup_{|f| \leq w} \beta_f(\mu)\]
	and the distributional CITE measure defined by
	\[\tilde{\beta}(\mu) \stackrel{\Delta}{=} \int_0^\infty \beta(\delta_x) \mu(dx).\]
	In view of the calculation of Section 3, specifically (\ref{integral}), it is evident that
	\[\beta_f(\delta_x) = \left| \int_0^\infty f(y) \frac{1}{r}m(\eta x,\,\eta y) dy \right|,\]
  where
	\[m(u,\,v) = 
	\begin{cases}
	1 - (u+v)e^{-v} & 0 \leq v \leq u,\\
	e^{-(v-u)} - (u+v)e^{-v}  & v > u.
	\end{cases}\]
	According to \cite{Pollard} p.60, 
	\[\beta(\delta_x) = \int_0^\infty \frac{1}{r}|m(\eta x,\,\eta y)| w(y)dy.\]
	Hence, if $w(y) = y^p$ for $p \geq 0$,
	\begin{eqnarray*}
	\beta(\delta_x)
	&=& \frac{1}{r} \eta^{-p-1} \int_0^\infty v^p |m(\eta x,\,v)|dv
	\end{eqnarray*}
	and
	\[\tilde{\beta}(\pi) = \frac{1}{r} \eta^{-p-1} \int_0^\infty \int_0^\infty
	v^p |m(u,\,v)| dv e^{-u} du.\]
	So, as in Section 3, we can analogously define the set of ``good states" as
	\begin{eqnarray*}
  \mathop{\mathscr H (r,\,\sigma^2;\,c)}
	&\stackrel{\Delta}{=}&	\{x:\, \beta(\delta_x) \leq c \tilde{\beta}(\pi)\}\\
	&=& EX(\infty) \mathop{\mathscr H (1,\,2;\,c)},
	\end{eqnarray*}
	so that $\mathop{\mathscr H (r,\,\sigma^2;\,c)}$ again scales similarly as does the
	set $\mathop{\mathscr G (r,\,\sigma^2;\,c)}$ of Section 3, due to the homogeneity of
	$w(y) = y^p$. (For the function $w(y) = e^{\theta y}$, such a scaling relationship
	does not hold.)
	
	By numerically computing $\tilde{\beta}(\pi)$ (via a numerical integration of $\beta(\delta_x)$
	against $\pi$), the graph of the set
	$\mathscr H (1,\,2) \stackrel{\Delta}{=}
	\{(x,\,c):\, x \in \mathop{\mathscr H (1,\,2;\,c)},\; c \geq 0\}$ 
	can be determined. In particular, it is computed below for $w(x) \equiv 1$; see Figure 2.
	Unlike the set $\mathscr G(1,\,2)$ of Section 3, it does not touch the $x$-axis.
	This is not surprising, because in the functional setting of Section 3, it will often
	be the case that the bias is monotone in the initial state. For example,
	for nondecreasing $f$, the stochastic monotonicity of RBM ensures that the bias
	will be negative for small values of $x$, and positive for large values of $x$, yielding
	(using a continuity argument) the existence of an intermediate $x$ at which
	$\beta_f(\delta_x)=0$. On the other hand, in the distributional setting, $\beta(\delta_x)$
	involves looking at the largest possible bias over a large class of functions $f$, and
	there typically will be no initial $x$ at which $\beta(\delta_x)$ will vanish.
	
	We further note that the shape of $\mathscr H(1,\,2)$ is quite different than that
	of $\mathscr{G}(1,2)$. For example, the minimizer of $\beta(\delta_x)$ as a function of 
	$x$ is achieved at a point
	$x^*$ that is smaller than that of $\beta_f(\delta_x)$ for $f(x) = x$. This occurs because
	$\beta(\delta_x)$ is defined in terms of the bias of bounded functions, while
	the set $\mathscr{G}(1,2)$ of Section 3 was computed for the identity mapping
	(which is unbounded). In computing the expectation of such a performance measure,
	it is advantageous to initialize the process at a larger value, because such an
	initialization will lead to higher likelihood paths that will quickly sample the large
	state values that typically contribute the most to the expectation of the performance
	measure. We further note that the set of good $x$'s associated with $\mathscr H(1,\,2)$ is a bit 
	smaller than that of Section 3 (in part, because $\mathscr H(1,\,2)$ involves a worst case bias,
	where as $\mathscr{G}(1,\,2)$ is determined only by a single function $f$). Nevertheless,
	the set of good states, with values of $\beta(\delta_x)$ of a magnitude less than
	or equal to that associated with initializing under the equilibrium distribution
	(i.e., $\tilde{\beta}(\pi)$), is large, and includes
	all the states $x$ with a value less than or equal to $1.84\, EX(\infty)$.
	\begin{figure}[H]
	\begin{center}
	\includegraphics[scale = 0.30]{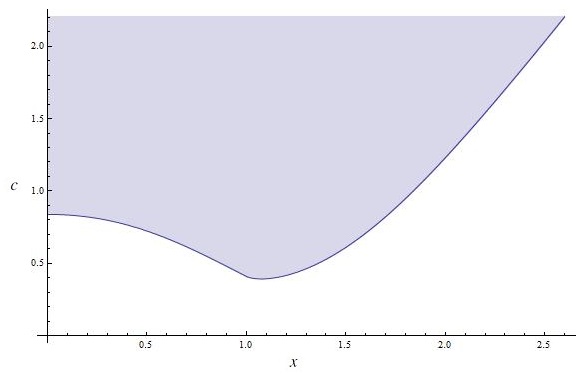}
	\caption{Plot of Good States: $EX(\infty)=1$}
	\end{center}
	\end{figure}

	\section{When Does Initial Transient Bias Matter?}

	Given the time-average estimator $\alpha(t) \stackrel{\Delta}{=} t^{-1} \int_0^t f(X(s))ds$
	for $\alpha = Ef(X(\infty))$, its (single replication) rate of convergence is determined by
	the central limit theorem (CLT). To develop the CLT for $\alpha(t)$, recall that
	$(M(t):\,t \geq 0)$
	is a martingale; see Section 3. The martingale CLT as applied 
	to the stochastic integral associated with the martingale then guarantees that
	\[\sqrt{t} (\alpha(t)-\alpha) \Rightarrow \kappa N(0,\,1)\]
  as $t \rightarrow \infty$, where $\kappa^2 = \sigma^2 Eh'(X(\infty))^2$;
	see, for example, p.339-340 of \cite{Ethier}. The CLT asserts that the
	expected stochastic variability of $\alpha(t)$,
	for large $t$, is governed by $\kappa E|N(0,\,1)| t^{-1/2}
	= \kappa (2/\pi)^{1/2} t^{-1/2}$.
	
	On the other hand, the systematic error in $\alpha(t)$ (namely, the bias) was analyzed
	in Section 3 and determined to be $\beta_f(\delta_x)/t$ for $t$ large, assuming that
	the process was initialized at state $x$. It is clear that for $t$ large enough, the stochastic
	variability dominates the bias. Specifically, for $t \geq t^*(x) \stackrel{\Delta}{=}
	\beta_f(\delta_x)^2 \pi/(2\kappa^2)$, the bias contribution is
	smaller	than the error due to stochastic variability.
	
	To help put the quantity $t^*(x)$ in perspective, note that the CLT suggests that a relative
	error $\epsilon^* = 2\kappa (2/\pi)^{1/2}/(t^*(x)^{1/2} |Ef(X(\infty))|)$
	is achieved at a run-length	$t^*(x)$ (contributed in equal measures from
	stochastic variability and bias). In other words,
	\[\epsilon^*(x) \stackrel{\Delta}{=}
	\frac{2\kappa^2 }{|\beta_f(\delta_x)|\cdot |Ef(X(\infty))|} \frac{2}{\pi}\]
	is the relative precision at which a simulation designed to achieve such an error tolerance will
	have an error that is contributed equally from stochastic variability and initial transient bias.
	At all smaller values of the relative precision, the stochastic variability dominates the error.
	In particular, to achieve a relative error tolerance of $\epsilon = \nu \epsilon^*(x)$
	(with $\nu < 1$), the associated run-length $t$ that must be used is such that the error due
	to initial transient bias at such a run-length is roughly a proportion $\nu$ of the stochastic
	error. Thus, the quantity $\epsilon^*(x)$ is an important measure of the threshold error
	tolerance at which the initial transient is no longer a dominant source of error in computing
	steady-state quantities.
	
	Figure 3 below provides a graph of the \textit{threshold error tolerance}
	as a function of $x$ for $f(y)=y$, with
	$(r,\,\sigma^2)=(1,\,2)$. Given that a relative error precision of $10\%$ or less
	is typically desired, we note that the set of states $x$ for which $\epsilon^*(x) \geq 0.1$
	is large, specifically the interval $[0,\,10.19]$.
	\begin{figure}[H]
	\begin{center}
	\includegraphics[scale = 0.50]{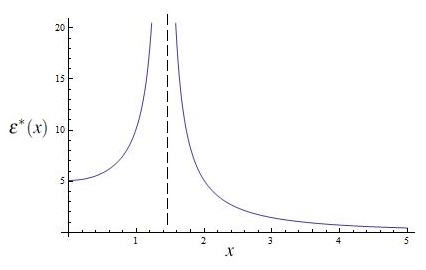}
	\caption{Plot of $\epsilon^*(x)$ vs. $x$: $r=1$ and $\sigma^2=2$}
	\end{center}
	\end{figure}

	\section{Mean Square Error for RBM Equilibrium Calculations}
	
	Because mean square error (MSE) is so frequently utilized in theoretical analyses of the
	initial transient problem, we provide here a detailed analysis of MSE in the setting of RBM.
	For a given performance measure $f$, recall the martingale $(M(t):\, t \geq 0)$ of Section 3.
	Hence, in view of the martingale property of the stochastic integral,
	\begin{eqnarray*}
	E_x\left( \int_0^t f_c(X(s))ds \right)^2
	&=& \sigma^2\int_0^t E_x h_c'(X(s))^2ds + h_c^2(x) + E_x h_c^2(X(t))\\
	&\qquad-& 2E_x h_c(x)h_c(X(t))
	- 2\sigma E_x h_c(X(t)) \int_0^t h_c'(X(s))dB(s).
	\end{eqnarray*}
	But
	\begin{eqnarray*}
	E_x h_c(X(t)) \int_0^t h_c'(X(s))dB(s)
	&=& E\left[ h_c(X(0))\int_{-t}^0 h_c'(X(s))dB(s)\,\bigg|\,X(-t)=x \right]\\
	&=& Eh_c(X^*(0))\int_{-\infty}^0 h_c'(X^*(s))dB(s) + o(1)
	\end{eqnarray*}
  as $t \rightarrow \infty$, where $o(1)$ is a deterministic function tending to
	$0$ as $t \rightarrow \infty$, and $(X^*,\,B) = ((X^*(t),\,B(t)):\,-\infty < t < \infty)$
	is such that $X^*$ is a stationary RBM driven by the Brownian motion $B$. Of course,
	\begin{equation}\label{inversion}
	Eh_c(X^*(0)) \int_{-\infty}^0 h'_c(X^*(s))dB(s) = -Eh_c(X^*(0)) \int_0^\infty
	h_c'(X^*(-s)) dB(-s).
	\end{equation}
	Since $X^*$ is a reversible one-dimensional diffusion driven by $B$ (see, for example,
	\cite{reversible}),
	\[((X^*(-t),\,-B(-t)): -\infty < t < \infty) \stackrel{\mathscr D}{=} ((X^*(t),\,B(t)):\,
	-\infty < t < \infty),\]
	where $\stackrel{\mathscr D}{=}$ denotes equality in distribution. Consequently,
	(\ref{inversion})	equals
	\[Eh_c(X^*(0)) \int_0^\infty h_c'(X^*(t)) dB(t) = 0,\]
	because of the martingale property of the stochastic integral.
	
	Furthermore, $E_x h_c^2(X(t)) = Eh_c^2(X(\infty)) + o(1)$ as $t \rightarrow \infty$
	and $E_x h_c(X(t)) = Eh_c(X(\infty))+o(1)$ as $t \rightarrow \infty$. Finally, by following
	the same argument as in Section 3, we find that
	\begin{eqnarray*}
	\int_0^t E_x h'(X(s))^2 ds
	&=& t Eh'(X(\infty))^2 + \int_0^t (E_x h_c'(X(s))^2-Eh_c'(X(\infty))^2)ds\\
	&=& \kappa^2 t/\sigma^2 + k_c(x) + o(1)
	\end{eqnarray*}
	as $t \rightarrow \infty$, where $k_c$ is the solution of the Poisson's equation associated
	with $h_c'^2$, namely $k_c$ satisfies
	\[(\mathscr L k_c)(x) = -(h_c'(x)^2-Eh_c'(X(\infty))^2),\;\;x \geq 0\]
	subject to $k_c'(0)=0,\;\;Ek_c(X(\infty)) = 0$.	In view of (\ref{integral}),
	\[k_c(x) = -\frac{1}{r}\int_0^x h_c'(y)^2 \left(e^{-\eta(y-x)}-1\right) dy
	+ \frac{2}{\sigma^2} \int_0^\infty \int_0^x h_c'(y)^2(e^{-\eta y}-e^{-\eta x})dydx.\]
	With $k_c$ now computed, the mean square error of $\alpha(t)$ is given by
	\begin{equation}\label{MSE}
	E(\alpha(t)-\alpha)^2 = \frac{\kappa^2}{t} + \frac{\sigma^2k_c(x)}{t^2} + \frac{h_c(x)^2}{t^2}
	+ \frac{Eh_c^2(X(\infty))}{t^2} + \frac{1}{t^2}o(1)
	\end{equation}
	as $t \rightarrow \infty$. In particular, for $f(x)=x$,
	\begin{eqnarray*}
	E(\alpha(t)-\alpha)^2
	&=& \frac{\sigma ^6}{2 r^4 t}
	+ \frac{\sigma ^2 \left(2 r^3 x^3+3 r^2 \sigma ^2 x^2-3 \sigma ^6\right)}{6 r^6 t^2}
	+ \frac{\left(\sigma ^4-2 r^2 x^2\right)^2}{16 r^6 t^2}
	+ \frac{5 \sigma ^8}{16 r^6 t^2} + \frac{1}{t^2}o(1)\\
	&=& \frac{\sigma ^6}{2 r^4 t}
	+ \frac{6 r^4 x^4+8 r^3 \sigma ^2 x^3+6 r^2 \sigma ^4 x^2-3 \sigma ^8}{24 r^6 t^2}
	+ \frac{1}{t^2}o(1)
	\end{eqnarray*}
	as $t \rightarrow \infty$.

	The term $h_c(x)^2/t^2$ is the squared bias contribution to the MSE due to the initial transient.
	The expression (\ref{MSE}) makes clear that the MSE includes other state-dependent
	contributions of the same order of magnitude (that are contributed by the variance of
	$\alpha(t)$ rather than the bias), namely $\sigma^2 k_c(x)/t^2$. Hence, a full analysis
	of the MSE impact of the effect of the initial transient should also (ideally) include
	an analysis of this variance term involving $k_c$, in addition to the bias contribution
	that is typically included in such an MSE analysis.

	\section{Conclusion}

	We have developed various formulae related to the initial transient problem for RBM.
	These formulae can be used directly in a simulation context, to approximate the impact of
	the initial transient for queueing simulations for which RBM is a suitable guide (eg.
	simulations of queues in heavy traffic). Our formulae also make clear a key insight that
	is likely true in a much broader class of simulations. In particular, for RBM, there is
	typically a robust set of initializing states for which the impact of the initial transient
	is roughly comparable to that associated with initializing in equilibrium. Since initializing
	in equilibrium corresponds to a setting in which there is no initial transient, this suggests
	that one should not worry excessively about the initial transient unless one has inadvertently
	initialized the simulation with a very poor (``bad") choice of state. If one instead
	initializes with a reasonable (``good") choice of state, the key element to a successful
	calculation of $Ef(X(\infty))$ is ensuring that the run-length is long enough to ensure
	that the stochastic variability has been reduced to a level commensurate with the desired
	accuracy. This suggests that a focus of initial transient research should be on building
	reliable algorithms for identifying settings in which one has inadvertently chosen a poor
	initialization that induces a large transient.

	\section{Acknowledgements}
	
	We thank the referees for their helpful comments and suggestions.
	Rob J. Wang is grateful to be supported by an Arvanitidis Stanford Graduate Fellowship
	in Memory of William K. Linvill, as well as an NSERC Postgraduate Scholarship (PGS D).


  \section{Author Biographies}
	
  ROB J. WANG is currently a Ph.D. candidate in Management Science and Engineering
  at Stanford University, specializing in Operations Research. He holds a B.Sc. (Honours) in 
  Mathematics from Queen's University in Kingston, Ontario (Canada). His research interests
  include simulation output analysis, queueing theory, diffusion processes, and statistical 
  inference. His email address is robjwang@stanford.edu and his
  webpage is http://web.stanford.edu/$\sim$robjwang/.\\

  \noindent
  PETER W. GLYNN is currently the Chair of the Department of Management Science and Engineering
  and Thomas Ford Professor of Engineering at Stanford University.
  He is a Fellow of INFORMS and of the Institute of Mathematical Statistics, has been
  co-winner of Best Publication Awards from the INFORMS Simulation Society in 1993 and 2008,
  and was the co-winner of the John von Neumann Theory Prize from INFORMS in 2010. In 2012, he
  was elected to the National Academy of Engineering. His research interests lie in stochastic   
	simulation, queueing theory, and statistical inference for stochastic processes.
	His email address is glynn@stanford.edu and his webpage is http://web.stanford.edu/$\sim$glynn/.

	\bibliographystyle{plainnat}
  \bibliography{WSCrefs}

  \end{document}